\documentclass[letterpaper, 10 pt, conference]{ieeeconf}  

\IEEEoverridecommandlockouts                            
\let\labelindent\relax
\usepackage{enumitem}
\usepackage{colorblock} 
\usepackage{hidecomments} 

\newcommand{\pfrac}[2]{\frac{\partial#1}{\partial #2}}

\def \vp {\varphi}

\newcommand{\remove}[1]{}

\title{\LARGE \bf Numerical Discretization Schemes that  \olive Preserve Flatness}

\author{Ashutosh Jindal$^{1}$, Florentina Nicolau$^{2}$, David Mart{\'\i}n Diego$^{3}$ and Ravi Banavar$^{4}$
\thanks{$^{1,4}$
        Systems and Control Engineering, Indian Institute of Technology Bombay, Mumbai, 400076, Maharashtra, India  $^1${\tt\small jindal.ashutosh21@gmail.com}, $^4${\tt\small banavar@iitb.ac.in} }
\thanks{$^{2}$  Quartz EA7393, École Nationale Supérieure de l'Électronique et de ses Applications, Cergy, France
{\tt\small florentina.nicolau@ensea.fr}}
\thanks{$^{3}$ Instituto de Ciencias Matem\'aticas (CSIC-UAM-UC3M-UCM), Calle Nicol\'as Cabrera 13-15, 28049 Madrid, Spain
{\tt\small david.martin@icmat.es}}
}
\usepackage{mymacros}
\usepackage{ragged2e} 
\usepackage{derivative}
\usepackage{tikz}
\usetikzlibrary{arrows.meta, decorations.pathreplacing}

\begin{document}
\maketitle
\thispagestyle{empty}


\begin{abstract}
Differential flatness serves as a powerful tool for controlling \olive{continuous time} nonlinear systems in problems such as motion planning and trajectory \olive{tracking}. A similar notion, called difference flatness, exists for discrete-time systems.
\olive{Although many control systems evolve in continuous time, control implementation is performed digitally, requiring discretization. It is well known in the literature that discretization does not necessarily preserve structural properties, and it has been established that, in general, flatness is not preserved under discretization (whether exact or approximate).
In this paper, inspired by our previous work  \cite{retr_disc_map} and based on the notion of discretization maps, we construct numerical schemes that preserve flatness.} 

\end{abstract}


\setlength{\parindent}{5pt}
\section{Introduction}

Differential flatness, \olive{introduced by Fliess, Lévine,} Martin, and Rouchon in the 1990s \cite{fliess1995flatness}, offers a powerful framework for the control of nonlinear systems. It has become a cornerstone in areas such as trajectory planning, optimal control, and feedback design, \olive{see, e.g.,} \cite{wang2022trajectory,martin:cel-00392180,kolar2017time}. \olive{For instance,} many mechanical systems, including quadrotors and gantry cranes, naturally exhibit flatness \cite{murray1995differential}, and this property has been extensively exploited in recent control applications \cite{mellinger2011minimum,greeff2018flatness,sun2022comparative,welde2022role}.
\olive{The fundamental property of flat systems is that all their trajectories can be parametrized by $m$ functions and their time derivatives, where $m$ denotes the number of control inputs.
More precisely, the continuous-time control system
$
\Sigma_c:~ \dot x = f(x,u),
$
where $x \in X \subset \mathbb{R}^n$ is the state and $u \in U \subset \mathbb{R}^m$ the control input,
is said to be \emph{flat} if there exist $m$ smooth functions $\varphi_i(x,u,\dots,u^{(p)})$, for some $p \geq 0$, such that
\vspace{-0.25cm}
\begin{equation} \label{descript}
x=\gamma(\varphi,\dots,\varphi^{(s-1)}), \qquad
u=\delta(\varphi,\dots,\varphi^{(s)}),\vspace{-0.25cm}
\end{equation}
for a certain integer $s$, where $\varphi=(\varphi{_1},\dots,\varphi{_m})$ denotes a \emph{flat output}.
Therefore, all state and control variables can be expressed in terms of the flat outputs without integration, and all trajectories of the system can be completely parametrized.}

The discrete-time analogue, known as difference flatness, extends this concept by relating the system variables to the \olive{flat output components} and their forward shifts \cite{kolar2022necessary}. For a detailed discussion and theoretical foundations of flatness, we refer the reader to \cite{fliess1995flatness,levine2009analysis,martin2001part1,martin2001part2} for continuous-time systems and \cite{diwold2021trajectory,kolar2016construction,kolar2022necessary,diwold2023theory} for their discrete-time counterparts. Similar to the continuous-time case, difference flatness has numerous applications in the design of control algorithms for discrete-time systems, \olive{see, e.g.,} \cite{diwold2021trajectory,diwold2022discrete,diwold2023discrete}.

\remove{For more information on flatness, we refer the reader to \cite{fliess1995flatness,fossas1998flatness, levine2009analysis, martin2001part1,martin2001part2,  sira2004differentially} for continuous time systems, and to \cite{diwold2021trajectory,kaldmae2013flatness,kolar2016construction,sira2004differentially,kolar2022necessary} for discrete time systems. For a comprehensive understanding of difference flatness, interested readers may also refer to the doctoral thesis of Johannes Diwold \cite{diwold2023theory}. } 
While most dynamical systems are continuous in nature, with the development of digital computation, the control algorithms are invariably implemented using digital controllers.  Such a digital controller consists of two components: (1) a sample-and-hold block that reads the system (and output) states at regular intervals and, between each interval, applies a constant control input; and (2) a numerical discretization block that numerically solves the continuous-time dynamics, which is then used to design control algorithms.

\remove{
While most dynamical systems are inherently continuous, the advent of digital computation has made it standard practice to implement control algorithms using digital controllers. Such controllers typically consist of two components: (1) a sample-and-hold block that measures system states (and outputs) at regular intervals and applies a constant control input between samples, and (2) a numerical discretization block that approximates the continuous-time dynamics, forming the basis for discrete-time control design.}

Although differential flatness provides a powerful framework for control design, \olive{this property is not, in general, preserved under sampling or numerical (or even exact) discretization.} For a given differentially flat continuous-time system, its discretization can often fail to be flat. This restricts the application of difference-flat methods. In \cite{diwold2023theory}, Diwold illustrates this with an example where the Euler discretization of a flat continuous-time system fails to remain flat in the discrete domain.
\olive{In fact, the preservation of flatness strongly depends}  on the chosen discretization scheme. In the same work, an alternative discretization method is proposed that maintains flatness. Since flatness greatly simplifies control design, preserving it under discretization is of significant interest. \olive{In particular,} Diwold \cite{diwold2023theory} further presents systematic constructions of both explicit and implicit Euler-based discretization schemes for a class of \olive{differentially} flat systems with a triangular structure. However, for general nonlinear systems, developing flatness preserving discretizations remains an open and challenging problem. \olive{To fully exploit discrete time flatness, it is thus essential to identify or design discretization schemes that retain this property.}

\remove{Although flatness is a powerful technique for designing control algorithms. As a system property, flatness is often not preserved under sampling and numerical (or even exact) discretization. For a given differentially flat continuous time system, its discretization can often fail to be flat. This restricts the application of difference flat methods. In \cite{diwold2023theory}, the authors with an example, show that for a given continuous time system, its Euler discretization fails to be flat in the discrete time. Furthermore, this is dependent upon the choice of (numerical) discretization technique used. In the same paper, the author presents an alternate discretization scheme that preserves flatness. Since flatness greatly simplifies control, it is of interest to preserve this property under discretization. In \cite{diwold2023theory}, the authors present a systematic way to construct Explicit and Implicit Euler based discretization schemes for a class of systems that are structurally flat and possess a triangular structure. However, for general systems, constructing a flatness preserving discretization remains an open problem. In order to utilize discrete time flatness, it is necessary to find discretization schemes that preserve flatness.} 

In our previous work \cite{retr_disc_map}, we presented a systematic way to construct discretization schemes that preserve \olive{the property of} static feedback linearizability.
\olive{Flat systems can be viewed as a generalization of static feedback linearizable systems. Specifically, they are linearizable via dynamic, invertible, and endogenous feedback (see, e.g., \cite{fliess1995flatness,levine2009analysis}). In this article, we explore the relationship between dynamic feedback linearizability and flatness, and, inspired by \cite{retr_disc_map}, we develop discretization schemes that preserve flatness.}

\textbf{Contribution:} In this article, we utilize discretization maps \cite{21MBLDMdD} to systematically construct  \olive{first order numerical discretization schemes for differentially flat continuous time  systems, ensuring that the resulting discrete time systems are also flat.
The motivation for using discretization maps is that standard Euclidean discretization techniques may fail to preserve the fundamental property that system states remain on the underlying manifold, which can render the discretization physically meaningless (for example, a rotation matrix may no longer satisfy the properties of a rotation matrix after discretization).
Discretization maps exploit the geometric structure of the manifold to construct discretizations that ensure the system states remain on the manifold at all discrete time steps, making them particularly suitable for flatness preserving schemes.
}


\remove{\section{Introduction}
Designing controls for nonlinear systems is always a challenging process. Unlike linear invariant systems, for which a set of well-established system theoretic methods, such as proportional-integral-derivative (PID) control, pole placement, and state feedback exist, there are no such standard control methods that work for all nonlinear systems. For nonlinear systems, the control schemes are to be designed case by case and in an ad hoc manner. Often, such nonlinear systems are classified into subclasses, and control algorithms are designed for such subclasses. One such class of nonlinear systems is the differentially flat system. Differential flatness was introduced in control theory by Fliess, Levine, Martin, and Rouchon around the 1990s \cite{fliess1995flatness}. Flatness as a property is exhibited by several mechanical systems, some of which are cataloged in \cite{murray1995differential}. Differential flatness greatly simplifies the design of control algorithms for such nonlinear systems and allows us to utilize simple and efficient linear techniques for trajectory planning\cite{wang2022trajectory}, feedback design \cite{martin:cel-00392180}, and optimal control problems \cite{kolar2017time}. Flatness allows converting a given dynamical system into a chain of integrators and thus greatly simplifying control \cite{martin:cel-00392180}. 
%
%
In recent works \cite{mellinger2011minimum,greeff2018flatness,sun2022comparative,welde2022role}, flatness-based ideas have been utilized to design control for quadrotors and other mechanical systems. The reason for the wide applicability of flatness based schemes, is due to their ability to render a dynamical problem (involving ordinary and partial differential euqations) to solving an algebraic problem. While we shall formally introduce flatness in the subsequent sections, informally, a dynamical system is called flat if there exist (flat) outputs such that the state and control for the system can be expressed as smooth mappings of the flat outputs and their derivatives. For more information on differential flatness, we refer the reader to look into \cite{fliess1995flatness,fossas1998flatness, levine2009analysis, martin2001part1,martin2001part2,  sira2004differentially} and references within.

For discrete time systems, a similar notion called difference flatness exists. A given discrete time system is called (difference) flat if there exists a flat output such that the system state and output can be described as smooth mappings of the forward shifts of the flat output \cite{kolar2022necessary}. In \cite{kolar2022necessary}, Kolar et al. provide a set of necessary and sufficient conditions under which a given discrete time system is flat. In recent years, difference flatness has been extensively studied by researchers, from which a few references are \cite{diwold2021trajectory,kaldmae2013flatness,kolar2016construction,sira2004differentially,kolar2022necessary}. For a detailed study on difference flatness and its applications, interested readers can refer 
to the doctoral thesis of Johannes Diwold \cite{diwold2023theory}. Similar to continuous time systems, difference flatness has several applications in designing control algorithms for discrete time systems for problems such as trajectory tracking, motion planning, etc.\cite{diwold2021trajectory,diwold2022discrete,diwold2023discrete}.

%
With the developments in digital computation, invariably all control algorithms are implemented using digital controllers. A continuous time system is implemented using a digital controller that consists of two components: (1) a sample and hold block that reads the system (and output) states at regular intervals and between each interval applies a constant control input; and (2) a numerical discretization block that numerically solves the continuous time dynamics, which is then used to design control algorithms. 
Like feedback linearization, flatness as a system property in general is not preserved under sampling and numerical (or even exact) discretization. In \cite{diwold2023theory}, authors show that, for a given differentially flat continuous time system, its Euler's discretization fails to be differentially flat. Furthermore, this property is dependent on the choice of discretization. In the same work \cite{diwold2023theory}, the authors present a way to construct Implicit and Explicit Euler based discretizations for a specific class of differentially flat systems that are structurally flat and possess a triangular structure. However, for general systems, constructing flatness preserving discretizations remains an open problem. 

\textbf{Contributions:} In our previous work \cite{retr_disc_map}, we present a systematic method to construct discretization schemes for static feedback linearizable systems that preserve feedback linearizability, which was then suitably extended to the dynamic feedback linearizable systems in this \cite{jindal2024constructing}. In this article, we utilize the connection between flatness and feedback linearization to design discretization schemes that preserve flatness.

\remove{\textbf{Notations:} \red{FN: Do we need this fisrt paragraph?} We use standard notations in our article. The set of real numbers is denoted by $
\R$, and the set of positive integers is denoted by $\N$. For any given set $\mathcal{A}$, and an integer $k\in\N$, $\mathcal{A}^k
\coloneqq \mathcal{A}\times \mathcal{A}\ldots \times \mathcal{A}$ is the $k$ times cartesian product. For any continuous signal $t\mapsto x(t)$, $\dot{x}(t)$ denotes the time derivative and $x^{(l)}\coloneqq\odv[l]{x}{t}$ is the $l$th order derivative. 

\red{FN: The following 2 paragraphs are not very precise. I would give all this later, in Section IIB, see my modifications.}

\blue{The notation $x(k)$ is adopted to denote the value of the variable $x$ at time $kh$, where $h$ is the discretization period. The operator $\fwshift$ is used to denote the forward shift  by one discretization
interval, i.e., $\fwshift x(k) = x(k+1)$, 
which we also denote by using  the superscript "$+$", that is, $x^{+} = x(k+1)$.} For any given $l>0$, $\fwshift[l]x =\fwshift[l-1](\fwshift x)$ is  a repeated application of the forward shift operator and denotes $\fwshift[l]x(k) = x(k+l)$. Let $x_1,x_2,\ldots,x_r$ be discrete time signals, then for any given smooth map $\psi(x_1,x_2,\ldots,x_r)$, $\fwshift[l]\psi(x_1,x_2,\ldots,x_l)$ is defined as $\psi(\fwshift[l]x_1,\fwshift[l]x_2,\ldots,\fwshift[l]x_r)$. 

For any continuously differentiable map $x \mapsto \psi(x)$, $\D\psi(x)$ denotes its jacobian, and for any
given multivariate map $(x,y,...,w)\mapsto F(x,y,...,w)$, $\D_xF(x,y...,w),...,\D_wF(x,y...,w)$, represent the partial derivatives of $F$ with respect to the respective arguments.}}

\section{Differential and Difference Flatness} \label{sec:flatness}
In this section, we \blue{recall} the notions of differential and difference flatness for continuous and discrete time systems. 

\subsection{Differential Flatness (Continuous Time)}\label{sec:flatness cont}

The definitions and associated results in this section are taken from standard sources on flatness, such as \cite{fliess1995flatness,fossas1998flatness, levine2009analysis, martin2001part1,martin2001part2,  sira2004differentially}.

Let $\stateset\subset\R[n]$ and $\controlset\subset\R[m]$ be open and nonempty subsets (or more generally be $n$ and $m$ dimensional manifolds, respectively), 
and consider the following nonlinear continous time control system on $\stateset$ defined by the differential equation
\begin{equation}
 \label{ctssys}
 \tag {$\Sigma_{\blue c}$}
 \dot{x}(t) = f(x(t),u(t)), 
\end{equation}
where $x(t)$ and $u(t)$ respectively denote the system state and control input at time $t$. 
The dynamics $f$ are such that Rank$(\D_u{f})= m$, where $\D_u{f} = \big(\pfrac{f_i}{u_j}\big)_{i,j}$, and are supposed $\mathcal{C}^{\infty}$-smooth, i.e., continuously differentiable up to any order.
Fix an integer $l \geq -1$ and denote $\controlset^l = \controlset \times \mathbb{R}^{ml}$ and $\bar u^l = (u,\dot  u, \dots, u^{(l)})$. For $l=-1$,  the set $ \controlset^{-1} $ is empty and $\bar u^{-1}$ in an empty sequence.

\begin{defn}[Differential Flatness]
The system \eqref{ctssys} is called \emph{differentially flat} at $(x_0, \bar u_0^l) \in X \times \controlset^l$, for  $l\geq -1$, if there exist 
{an open set $\mathcal O^{l}$ containing} $(x_0, \bar u_0^l)$
and~$m$ smooth functions $\vp_i=\vp_i(x, u,\dot  u, \dots, u^{(p)})$, {with $p\leq l$}, $1 \leq i\leq m$, defined in $\mathcal O^l$, having the following property: there exist an integer $s$ and smooth functions~$ \gamma_i$, $1 \leq i \leq n$, and~$\varsigma_j$, $ 1 \leq j \leq m$, such that
$$
x_i = \gamma_i (\vp, \dot \vp, \dots, \vp^{(s-1)}) \mbox{ and } u_j=  \varsigma_j (\vp, \dot \vp, \dots, \vp^{(s)})
$$
for any $C^{l+s}$- control $u(t)$ and corresponding trajectory~$x(t)$
that satisfy $(x(t),$ $ u(t), \dots ,$ $u^{(l)}(t)) \in~\mathcal O^l$, where $\vp=(\vp_1, \dots, \vp_m)$ and is  called a \emph{flat output}.
\end{defn}

The notion of differential flatness is closely related to that of feedback linearization.
Indeed, it has been established in the literature, see \cite{fliess1995flatness, martin1992phd,pomet1995differential}, that the class of  differential flat systems is equivalent to \blue{that} of systems linearizable by dynamic, invertible and endogenous feedback.
More precisely, the continuous time control system \eqref{ctssys} is flat if and only if there exist a dynamic endogenous precompensator
\begin{equation*}
\dot{y} = g(x,y,v)~\quad y\in\compstateset\subset\R[r], ~v\in\compcontrolset\subset\R[m],
\end{equation*}
and an invertible feedback $u= \psi(x,y,v)\in\R[m]$ such that the extended system
\begin{equation}
\label{preconsys}
\tag{$\Pi$}
    \dot{x} = f(x,\psi(x,y,v)),~
    \dot{y}= g(x,y,v),
\end{equation}
can be brought via a local change of coordinates\footnote{\olive{The static feedback transformation that may à priori be applied on~\eqref{preconsys} for (static) feedback linearizability, see \cite{l1980linearization}, is contained in $\psi$.}} $\linstate = \extdiff(x,y)$ 
 %
%
into a linear controllable system of the form 
\begin{equation}
\label{lin_sys}
\tag{$\Lambda$}
\dot{\linstate} = \mathbf{A}\linstate +\mathbf{B}v,
\end{equation}
where $\mathbf{A}$ and $\mathbf{B}$ are matrices corresponding to the Burnovsky Controllable Canonical Form\footnote{Since every linear system can be converted into a canonical form by a linear transformation, such a form always exists and can be readily obtained.}.

The above precompensator is said to be endogenous if there exist a smooth function $\mu$ and an integer $\ell$ such that  $y= \mu(x,u,\dot{u},\ldots,u^{(\ell)})$; thus the precompensator does not introduce new states in the system, $y$ being generated by the original state $x$, original control $u$ and its successive time derivatives.
The feedback is invertible if one can uniquely express $v$ as $v = \bar\nu(x,y,u,\ldots,u^{(\ell)})$, which in the endogenous case simplifies to
\begin{equation}
    \label{dzncont}
    v = \bar\nu(x,y,u,\ldots,u^{(\ell)})= \nu (x,u,\ldots,u^{(\ell)}).
\end{equation}
\remove{
\begin{defn}{D-equivalence}
Consider a control system 
\begin{equation}
\label{ctsys2}
\tag{$\tilde\Sigma$} 
\dot{\tilde x} = \tilde f(\tilde x,\tilde u),~\quad \tilde x\in\tilde\stateset\subset \R[\tilde n], \tilde u\in\tilde\controlset\subset \R[\tilde m] 
\end{equation} 
Then \eqref{ctssys} and $\eqref{ctsys2}$ are $\mathrm{D}$-equivalent if there exists two integers $l$ and $\tilde{l}$ and two pair of maps
$(\phi,\psi):\stateset^{l}\times\controlset\lra\tilde\stateset\times\tilde\controlset$ and $(\tilde\phi,\tilde\psi):\tilde\stateset^{l}\times\tilde\controlset\lra\stateset\times\controlset$ that map trajectories onto trajectories and are mutually inverse on trajectories.  
\end{defn}}
\remove{
\begin{thm}[Differential Flatness \cite{}]
A given dynamical system \eqref{ctssys} is flat if there exists a dynamic endogenous invertible feedback, such that the extended system \eqref{preconsys} is equivalent to a linear time-invariant system by a diffeomorphic change of coordinates 
\end{thm}
}
\remove{
\begin{thm}[\cite{fliess1995flatness}]
Two given dynamical system \eqref{ctssys} and \eqref{ctsys2} are $\mathrm{D}$ equivalent if there exists  endogenous and invertible precompensators $(\Pi)$ and ($\tilde\Pi$) such that the precompensated system ($\Sigma\circ\Pi$) and ($\tilde\Sigma\circ\tilde\Pi$) are equivalent through a diffeomorphic change of coordinates.
\end{thm}}
\remove{
\begin{defn}[Flatness (alternate definition)]
A given dynamical system \eqref{ctssys} is flat if and only if it is $\mathrm{D}$-equivalent to a linear system 
\begin{equation}
\label{linsys}
\tag{$\Lambda$}
\dot{z} = Az+Bv
\end{equation}
and inturn is $\mathrm{D}$-equivalent to a trivial system with no dynamics. 
\end{defn}
}
\subsection{Difference Flatness (Discrete time)}
Consider a discrete time control system given by 
\begin{equation}
\label{dtssys}
\tag{$\Sigma_d$}
x^{+} = \blue f(x,u)
\end{equation}
with $x\in\stateset\subset\R[n],u\in\controlset\subset\R[m]$, \blue{where the superscript "$+$" denotes the forward shift  by one discretization period}, and $\blue f:\stateset\times\controlset\lra\stateset$ is a $\mathcal{C}^\infty$ smooth map
\blue{such that 
Rank$(\D_{(x,u)}f) = n$.} 
%
%

This assumption is readily satisfied by discrete time systems arising from the discretization of a continuous time system, as for such systems one has
\blue{Rank$(\D_{x}f)= n$.  If~\eqref{dtssys} is the discretization of a continuous time system with discretization period $h$, the notation $x[k]$ is adopted to denote the value of the variable $x$ at time $kh$.}


\blue{For discrete time systems, difference flatness can be defined in a direct analogy with the continuous time case by replacing time derivatives with forward shifts, see for instance \cite{kaldmae2013flatness,kolar2016construction,kolar2022necessary, sira2004differentially}. Under this definition\footnote{There exists another definition allowing also for backward shifts, see, for instance, \cite{guillot2019flatness}, but we do not consider it here since in the terminology of \cite{aranda2008linearization}, such systems would be linearizable by an exogenous dynamic feedback.}, as in the continuous time case, flatness is equivalent to linearizability via an endogenous dynamic feedback, as defined for discrete time systems in \cite{aranda2008linearization}.}


%

\blue{Similarly to the continous-time case, the results on difference flatness are local; therefore, to preserve their local nature, we must ensure that 
\( x^+ \), as determined by~\eqref{dtssys}, remains sufficiently close to \( x \). 
The continuity of \( f \) guarantees that this can be achieved by restricting the analysis 
to a sufficiently small neighborhood of an equilibrium point \((x_0, u_0)\) of~\eqref{dtssys}, 
that is, a point \((x_0, u_0)\) satisfying \( f(x_0, u_0) = x_0 \). 
This assumption is standard in the discrete time control literature (see, e.g., \cite{grizzle1986feedback,kottainversion}).
}



\blue{Below, the operator $\fwshift$ is used to denote the forward shift  by one discretization period, i.e., $\fwshift x[k] = x[k+1]$, 
which we also denote by using  the superscript "$+$". 
In the definition of difference flatness, we also need higher forward shifts (associated to the control input and to the flat output): for fixed $l>0$, $\fwshift[l]u =\fwshift[l-1](\fwshift u)$ is  a repeated application of the forward shift operator and denotes $\fwshift[l]u[k] = u[k+l]$. For any given smooth map $\psi(x, u ,\fwshift u, \ldots,\fwshift[l]u)$,
the forward shift operator acts according to 
$\fwshift  \psi(x, u ,\fwshift u, \ldots,\fwshift[l]u) = \psi(f(x,u),\fwshift u,\fwshift[2]u, \ldots,\fwshift[l+1]u)$. 
}

\remove{Using the forward shift notation, system \eqref{dtssys} can be expressed as $\bwshift(x)\coloneqq x^{(1)} = F(x,u)$.}
\begin{defn}[Difference Flatness]
The discrete time system \eqref{dtssys} is difference flat \blue{around an equilibrium $(x_0,u_0)$} 
if there exist an open set $\mathcal{O}^{l}$ containing $(x_0,\bar u_0^{l}) = (x_0, u_0,\ldots,u_0)$ and $m$ smooth functions $\varphi_i = \varphi_i(x,u,\bwshift u,\ldots,\bwshift^{p}u)$ with $p\leq l$, $1\leq i\leq m$,  defined in $\mathcal O^l$, having the following property: there exist an integer $s$ and smooth functions $\gamma_i$, $1\leq i\leq n$ and $\xi_j$, $1\leq j\leq m$, such that 
\begin{equation*}
    x_i = \gamma_i(\varphi,\fwshift\varphi,\ldots,\fwshift[s-1]\varphi),
\text{ and }
    u_j = \xi_j(\varphi,\fwshift\varphi,\ldots,\fwshift[s]\varphi),
\end{equation*}     
for any control $k\mapsto u[k]$ and corresponding trajectory $k\mapsto x[k]$ satisfying $(x[k],u[k],\fwshift u[k], \ldots \fwshift[l]u[k])\in\mathcal{O}^{l}$, where $\varphi = (\varphi_1,\ldots,\varphi_m)$ is called a \emph{flat output}. 

\remove{such that, $(x,u)\in\stateset\times\controlset$ can be represented as a smooth function of $y$ and their forward shifts $y^+,\ldots,\bwshift^q y$ up to some finite order $q$ i.e.,} 
\end{defn}

\remove{Similar to continuous time, a dynamical precompensator in the discrete time can be defined as follows
\begin{defn}[Discrete time dynamical feedback]
Consider the discrete time system \eqref{dtssys}, together with a dynamic compensator
\begin{equation}
\label{dtsprecon}
\tag{$\Pi_d$}
 y^{+} = G(x,y,v), \quad y\in\compstateset\subset\R[r], ~v\in\compcontrolset\subset\R[m]
\end{equation}
and a feedback $u=\Psi(x,y,v)\in\controlset$ such that the extended system becomes
\begin{equation}
\label{dtspreconsys}
\tag{$\Sigma_d\circ\Pi_d$}
x^{+} = F(x,\psi(x,y,v)),~ y^{+} = G(x,y,v)
\end{equation}
The precompensation is said to be endogenous if there exists a smooth function $\Omega$ such that $y = \omega(x,u,u^+,\ldots,\fwshift[l]u)$, which depends only upon the forward shifts of the input variable $u$ up to some finite $l$ and is invertible if one can uniquely express $v$ as
$v= \nu(x,u,\ldots,u^{(l)})$
\end{defn}}

\section{Flatness Preserving Discretization}

It has been established in the literature that, in general, the flatness is not preserved under discretization (whether exact or approximate). 
\blue{For example, Grizzle \cite{grizzle1988feedback} demonstrates that static feedback linearizability (a property satisfied by a special subclass of flat systems, see Section~\ref{sec:flatness}) is not preserved under discretization.} 
\begin{example}
To illustrate this phenomenon, consider the following example from \cite{diwold2023theory}:
\begin{equation}
 \begin{split}
 \dot{x}_1 = x_2^2+x_2+x_1u,~
 \dot{x}_2 = (1+x_2)u
 \end{split}  
\end{equation}
\blue{which} is flat with flat output $y=x_1/(1+x_2)$. However, \blue{its} Explicit Euler discretization 
\begin{equation*}
 {x}_1^{+} = x_1+h(x_2^2+x_2+x_1u),~{x}_2^{+} = x_2+h(1+x_2)u
\end{equation*}
where $h>0$ is the \blue{discretization period}, is not flat for any choice of $h>0$. \hfill $\triangleright$
\end{example}


\blue{In \cite{retr_disc_map}, we have demonstrated that for a given static feedback linearizable continuous time system, using discretization maps \cite{21MBLDMdD}, it is possible to construct discretization schemes that preserve feedback linearizability.} 
\magenta{Inspired by this work, the goal of the current paper is to show that a similar result can be used for generating flatness preserving discretizations schemes.}

\subsection{Discretization Maps}
%
%
\olive{When discretizing,} \blue{Euclidean techniques ensure that the evolution of the ensuing discretized system \olive{occurs} on a
Euclidean space. Apart from this basic requirement, different discretization schemes ensure
different requirements in terms of the deviation from the original nonlinear system.
However, for systems evolving on nonlinear manifolds, such Euclidean schemes do not 
\olive{even} 
guarantee  the basic requirement 
that the system state remains on the manifold for all time instants. \cyan{This deprives
the {discretization of its relevance since the values of the discretized variables} lose their 
physical significance in the context of the problem.
For instance, a $3 \times 3$ matrix of real numbers may
no longer sastify the \olive{properties} of a $3 \times 3$ rotation matrix ($\det(\cdot)=1$ and $(\cdot)^\top (\cdot) = I$, \olive{where~$I$ denotes the identity matrix})}.}

%

\blue{Discretization maps, as we refer to in this article, are a class of maps that utilize the geometric properties of the manifold to construct discretizations such that the system state remains on the manifold for all $k\in \N$, where $k$ is the iterating index of the discretized trajectory.}
\begin{defn}[Discretization Maps \cite{21MBLDMdD}]
Let $\manifoldM$ be a smooth manifold and $\T\manifoldM$ be the associated tangent bundle. Let $\mathcal{O}\subset \T\manifoldM$ be   an open neighborhood of the zero section of the tangent bundle $\T\manifoldM$. Then $\discmap(x,\nu)\coloneqq (\discmap_1(x,\nu),\discmap_2(x,\nu))\in \manifoldM\times \manifoldM~\forall  (x,\nu) \in \mathcal{O}$
is a discretization map if, for every $x\in \stateset$, it satisfies 
\begin{enumerate}[label=\textup{(\roman*)},leftmargin=*, widest=b, align=left]
    \item $(x,0_x)\mapsto \discmap(x,0_x) = (x,x)$, and
    \item $\T_{(x,0_x)}\discmap_2-\T_{(x,0_x)}\discmap_1 \colon \T_{(x,0_x)}\T_x\manifoldM\simeq \T_x\manifoldM \lra T_x\manifoldM$ is the identity map on $T_x\manifoldM$,
\end{enumerate}
where $\T_{(x,0_x)}\discmap_i$ is the tangent map of $\discmap_i$, $i\in\{1,2\}$ at $0_x$ (the zero element in $\T_x\manifoldM$) and $\T_{(x,0_x)}\T_x\manifoldM$ is canonically identified with $\T_x\manifoldM$.     
\end{defn}
\begin{figure}
\centering
\resizebox{2.4in}{1.2in}{
    \begin{tikzpicture}
        \draw[dotted,white] (0,0) grid (9,5);
        \draw[gray,fill=gray,fill opacity=0.1] (1,3) to[out=15,in=105] (3,1) to[out=-5,in=-175] (7,1) to[out=105,in=15] (5,3) to[out=-175,in=-5] (1,3);
        \node at (2,1.5) {$\manifoldM$};
        \draw[fill] (4,2.5) circle (1pt);
        \draw[] (4,2.5) to[out=-30,in=110] (5.2,1);
        \draw[] (4,2.5) to[out=0,in=160] (6.4,2);
        \draw[fill] (4.65,2) circle (1pt);
        \draw[fill] (5,2.42) circle (1pt);
        \node at (3.7,2.5) {$x$};
        \node at (5.5,2.7) {$\discmap_2(x,v)$};
        \node at (3.8,2) {$\discmap_1(x,v)$};
        \node at (5,0.6) {$\discmap_1(x,tv)$};
        \node at (7.5,2) {$\discmap_2(x,tv)$};
        \draw[dotted] (4,2.5)--(4,4);
        \draw[fill=gray,fill opacity=0.4] (3.5,3.5)--(6,3.5)--(5,4.5)--(2.5,4.5)--cycle;
        \draw[fill] (4,4) circle (1pt);
        \draw[thick,-stealth] (4,4)--(4.8,3.7);
        \node at (3.7,4) {$0$};
        \node at (5,3.7) {$v$};
        \node at (2.5,3.5) {$\T_x\manifoldM$};
\end{tikzpicture}}
\vspace{-0.1in}
\caption{\magenta Discretization map $\discmap$ mapping points from $\T\manifoldM$ on to $\manifoldM\times\manifoldM$.}
    \label{discmap_scheme}
\end{figure}

Intuitively, for each point $x\in\manifoldM$ and a velocity vector $\nu\in\T_x\manifoldM$, the discretization map generates two points $\discmap_1(x,\nu)$ and $\discmap_2(x,\nu)$ on the manifold, and it preserves the velocities arbitrarily close to zero velocity $0_x$.
\olive{These two points represent the initial and next states after a small time step in a discretization scheme.
Condition (i) states that if the velocity is zero, the map returns the same point twice, meaning that when the system does not move, the “initial” and “final” states of the discrete time step coincide.
The intuition behind condition (ii) is that for small velocities close to zero, the difference between the two points, 
 exactly reproduces the initial tangent vector.
In other words, the discretization scheme locally preserves the velocity, ensuring that the discrete approximation closely matches the continuous time dynamics for small steps.
This defines a geometry preserving discretization scheme, which keeps the points on the manifold rather than allowing them to leave it.}

\maroon{A schematic representation of how discretization maps work is shown in Fig. \ref{discmap_scheme}}. \cyan{From the definition of the discretization map, it is possible to deduce that ${\mathcal D}$ is a local diffeomorphism from a tubular neighborhood of the zero section to a tubular neighborhood of the set of identities (points of the form $(x, x)$).} \olive{This ensures that the map can be locally inverted to recover the initial tangent vector from the two discrete points.}

\olive{To sum up,} discretization maps allow us to naturally construct numerical discretization for dynamical systems that respect the geometrical constraints and keep system states on the underlying manifold.

\begin{prop}[Discretization of \blue a vector field \cite{21MBLDMdD}]
\label{disc-scheme}
Let $\discmap$ be any arbitrary discretization map on $\T\stateset$ and \maroon{$\discmap^{-1}$ be its inverse}. Let $\dot{x} = f(x,u)$, $x\in\stateset$, $u\in\controlset$, be a given control system.  For a given stepsize $h>0$, and \maroon{a given control sequence $k\mapsto u[k]$}, define a sequence $k\mapsto x[k]$ using the following recursion
\begin{equation}
\label{recursion_disc_1}
    \begin{array}{l@{\,}l}
     \discmap'^{-1}(x,x^+) = &\left(\tau_\stateset(\discmap^{-1}(x,x^+)),\right. \\
     & \qquad \left. hf(\tau_{\stateset}(\discmap^{-1}(x,x^+)),u)\right) 
\end{array}
\end{equation}
with $\T\stateset\ni(x,\nu)\mapsto\tau_{\stateset}(x,\nu) = x$ as the canonical projection of $\T\stateset$ on to $\stateset$. 
Then \eqref{recursion_disc_1} results in a numerical discretization of $\dot{x}=f(x,u)$ with the zero order hold control {$u(t)= u[k]$ for all $t\in[kh,(k+1)h[$}, accurate at least up to first order\footnote{Consider a control system  $\dot{x} = f(x,u)$ referred ($\Sigma$) and for a given control input $u(t)$, let {$x(t)$}  be its exact solution. 
A numerical approximation $x^+= f_h(x,u)$ for ($\Sigma$) is called of order $r$ if there exist $K>0$ and some $h_0>0$ such that for all $0<h<h_0$, and $k\in\N$, $\norm{x((k+1)h)-f_h({x}(kh)),{u(kh)}))}/h\leq Kh^r$   (\magenta{see} \cite{blanes2017concise}).

It is possible to construct discretization maps that result in higher order discretization schemes, see \cite{21MBLDMdD}.
%
%
}.
\end{prop}

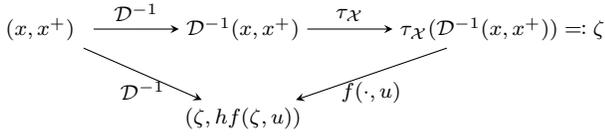
\begin{figure}
    \centering
    \begin{tikzpicture}
  \footnotesize{\matrix (m) [matrix of math nodes,row sep=2.5em,column sep=4em,minimum width=5em]
  { (x,x^+) & \discmap^{-1}(x,x^+) & \tau_\stateset(\discmap^{-1}(x,x^+))\eqqcolon\vectortemp \\
&(\vectortemp,hf(\vectortemp,u))&  \\};}
  \footnotesize{\path[-stealth]
    (m-1-1) edge node [below] {$\discmap^{-1}$} (m-2-2)
    (m-1-1)edge node [above] {$\discmap^{-1}$} (m-1-2)
    (m-1-2) edge node [above] {$\tau_\stateset$}
            (m-1-3)
    (m-1-3) edge node [below] {$f(\cdot,u)$} (m-2-2);}
\end{tikzpicture}
\vspace{-0.1in}
    \caption{\magenta Discretization scheme for $\dot x = f(x,u)$, from a given discretization map $\discmap$.}
\label{fig:desc_scheme_diag}
\end{figure}

The process of obtaining a discretization scheme from a given discretization map is represented by a schematic in Fig. \ref{fig:desc_scheme_diag}

%
\begin{example}
\label{disc_example}
\magenta{Let $\stateset =\R[n]$ and fix $\alpha\in[0,1]$. Then $\discmap(x,\nu)= (x-\alpha \nu,x+(1-\alpha)\nu)$, where $(x,\nu)\in \T\stateset$,} 
is a discretization map which results in the discretization scheme: $x^{+} = x +hf\big((1-\alpha)x^++\alpha x,u\big)$, 
which can be implicitly solved for $x^{+}$ to obtain a discrete system of type \eqref{dtssys}. \hfill $\triangleright$ 
\end{example}
 For each choice of discretization map $\discmap$, \magenta{relation} \eqref{recursion_disc_1} results in a different numerical discretization scheme for $\dot{x} = f(x,u)$. The task of finding flatness preserving discretization  \olive{scheme} then boils down to finding an appropriate discretization map.  In this direction, \magenta{we recall the} \blue{following result from \cite{retr_disc_map} that allows us to induce discretization between manifolds (see Fig.~\ref{fig:my_label}).} 



%
%

\remove{\florentina{Which reference is correct?
\\
$\discmap$  is define on $T\manifoldM$. We say several times discretization map $\discmap$ on $\manifoldM$. what formulation is standard ? discretization map $\discmap$ on $\manifoldM$ or discretization map $\discmap$ on $T\manifoldM$? -- corrected
}}
\begin{prop}[
{\cite{retr_disc_map}}]
\label{r_lift}
Let $\manifoldM$ and $\manifoldN$ be two manifolds such that $\manifoldM$ is diffeomorphic to $\manifoldN$. Let $\manifoldM\ni x\mapsto\phi(x)\in\manifoldN$ be the required diffeomorphism. Then for a given {discretization map $\discmap$ on $\T\manifoldM$}, $\discmap' \coloneqq\red{(\phi\times\phi)}\circ \discmap\circ \T\phi^{-1}$, \maroon{where $(\phi\times\phi)(x,x')\coloneqq (\phi(x),\phi(x'))\in\manifoldN\times\manifoldN$} and $(x,\nu)\mapsto\T\phi(x,\nu)\coloneqq (\phi(x),\D\phi(x)\nu)$ is the tangent lift of~$\phi$, is a  discretization map on $\T\manifoldN$  and viceversa. 
\end{prop}
\begin{figure}
    \centering
    \begin{tikzpicture}
  \footnotesize{\matrix (m) [matrix of math nodes,row sep=2.5em,column sep=3.5em,minimum width=1.5em]
  {
   \T \manifoldM & \T \manifoldN \\
    \manifoldM\times \manifoldM & \manifoldN\times \manifoldN \\};}
 \footnotesize{   
  \path[-stealth]
    (m-1-1) edge node [left] {$\discmap$} (m-2-1)
            edge node [above] {$\T\phi$} (m-1-2)
    (m-2-1.east|-m-2-2) edge node [below] {$\phi\times\phi$}
            (m-2-2)
    (m-1-2) edge node [right] {$\discmap'$} (m-2-2);}
\end{tikzpicture}
\vspace{-0.1in}
    \caption{$\discmap$ and $\discmap'$ commute\magenta.}
    \label{fig:my_label}
\end{figure}
  %
%
\subsection{Main Result: Flatness Preserving Discretizations}
Propositions \ref{disc-scheme} and \ref{r_lift} lie at the crux of the machinery to construct flatness preserving discretizations. Since the class of flat systems is equivalent to that of systems linearizable by dynamic endogenous feedback, the idea is to construct discretization maps on the \magenta{precompensated system brought into its linear equivalent}
and lift it using a diffeomorphism to construct discretization schemes for the original systems.

\magenta{Our main result is given by the following theorem constructing flatness preserving discretization schemes.}
\olive{Next}, for ease of notation, \olive{for a precompensated system \eqref{preconsys},} we define $\extendedstate\coloneqq(x,y)$ and compactly denote \eqref{preconsys} as $\dot{\extendedstate} = \mathbf{F}(\extendedstate,v)$, with $\mathbf{F}(\extendedstate,v)\coloneqq (f(x,y,v),g(x,y,v))$.

\begin{thm}[Flatness Preserving Discretization]
\label{main_result}
\magenta{Consider a flat continous time system \eqref{ctssys} and construct a precompensated system \eqref{preconsys}} \olive{that is locally diffeomorphic to a linear system \eqref{lin_sys} through a local change of coordinates~$\extdiff$, see Section~\ref{sec:flatness cont}}.
Let $\manifoldM\subset\stateset\times\compstateset$ be the local neighborhood on which $\extdiff$ is defined and $\manifoldN \coloneqq \extdiff(\manifoldM)$ be its image.
Let $\mathcal{O}'\subset\T\manifoldN$ be an open neighborhood of the zero section of~$\T\manifoldN$, \olive{the tangent bundle of $\manifoldN$, and} $\discmap':\mathcal{O}'\lra\manifoldN\times\manifoldN$ be a discretization map on $\olive \T\manifoldN$.
\olive{Then the following statements hold:}
\begin{enumerate}[label=\textup{(\roman*)},leftmargin=*, widest=b, align=left]
 \item \olive{The map
 \label{item1}
 \begin{equation}
\label{eq3.3}
\begin{array}{l@{\,}l}
     \discmap'^{-1}(z,z^+) = &\left(\tau_\manifoldN(\discmap'^{-1}(z,z^+)),\right. \\
     & ~~ \left. h(\mathbf{A}\tau_\manifoldN(\discmap'^{-1}(z,z^+))+\mathbf{B}v\right)
\end{array}
\end{equation}
defines the following linear  first order discretization of~\eqref{lin_sys}}
\begin{equation}
\label{dis_lin_sys}
\tag{$\Lambda_h$}
z^{+} = \mathbf{A}_hz+\mathbf{B}_hv,
\end{equation}
\maroon{where $\mathbf{A}_h$ and $\mathbf{B}_h$ are fixed matrices obtained by solving \eqref{eq3.3} for $\linstate^+$. }

\item \olive{The map}
\label{item2}
\begin{equation}
\label{linearizingdisc}
\discmap = (\extdiff^{-1}\times\extdiff^{-1})\circ\discmap'\circ\T\extdiff
\end{equation}
is a discretization map on $\olive \T\manifoldM$, and
\remove{
\begin{equation}
    \discmap^{-1}(\extendedstate,\extendedstate^+) = \big(\tau_{\manifoldM}(\discmap^{-1}(\extendedstate,\extendedstate^+)),h\mathbf{F}(\tau_{\manifoldM}(\discmap^{-1}(\extendedstate,\extendedstate^+)),v)\big)
\end{equation}}
\begin{equation}
 \label{recursion_disc}
\begin{array}{l@{\,}l}
     \discmap^{-1}(\extendedstate,\extendedstate^+) = &\left(\tau_\manifoldM(\discmap^{-1}(\extendedstate,\extendedstate^+)),\right. \\
     & \qquad \left. h\mathbf{F}(\tau_{\manifoldM}(\discmap^{-1}(\extendedstate,\extendedstate^+)),v)\right)
\end{array}
\end{equation}
is a first order discretization
of \eqref{preconsys}, which when solved for $\extendedstate^+$ and written as
\begin{equation}
\label{dis_ext_sys}
\tag{$\Pi_h$}
\olive \extendedstate^+ =  \pmat{x^+\\y^+}=\mathbf{F}_h(\extendedstate,v) = \pmat{f_h(x,y,v)\\g_h(x,y,v)}
\end{equation}
is diffeomorphic to \eqref{dis_lin_sys} through \olive{the} local coordinate change $z\coloneqq\extdiff(\extendedstate)$, and is therefore flat.
\end{enumerate}

\end{thm}

\smallskip

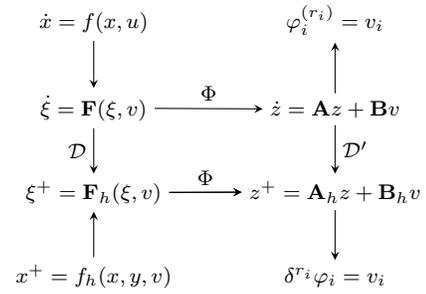
\begin{figure}
    \centering
    \begin{tikzpicture}
  \footnotesize{\matrix (m) [matrix of math nodes,row sep=2.2em,column sep=3em,minimum width=1.5em]
{
\dot{x}= f(x,u)& \maroon{\varphi_i^{(r_i)} = v_i}\\\dot{\extendedstate}= \mathbf{F}(\extendedstate,v)&  \dot\linstate = \mathbf{A}\linstate+\mathbf{B}v\\       \extendedstate^{+}=\mathbf{F}_h(\extendedstate,v)& \linstate^{+}= \mathbf{A}_h\linstate+\mathbf{B}_hv \\ x^+ = f_h(x,y,v)&\maroon{\fwshift[r_i]\varphi_i=v_i}\\};}
  \footnotesize{\path[-stealth]
    (m-2-1) edge node [left] {$\discmap$} (m-3-1)
    (m-2-1) edge node [above] {$\extdiff$} (m-2-2);
    \path[-stealth]
    (m-3-1) edge node [above] {$\extdiff$} (m-3-2)
    (m-2-2) edge node [below] {} (m-1-2)
    (m-3-1) edge node [below] {} (m-3-2)
    (m-3-2) edge node [below] {} (m-4-2)
    (m-2-2) edge node [right] {$\discmap'$} (m-3-2);
    \path[-stealth]
    (m-4-1) edge node [left] {} (m-3-1)
        (m-1-1) edge node [left] {} (m-2-1);
}
\end{tikzpicture}
\vspace{-0.1in}
    \caption{\olive Discrete time implementation of \eqref{ctssys} via a flatness preserving discretization scheme, with flat outputs $\varphi_i$, $1\leq i\leq m$.} 
    \label{fig:my_label2}
\end{figure}

\begin{proof}
\olive{Proof of \ref{item1} : Applying Proposition \ref{disc-scheme} \olive{to} \eqref{lin_sys}, 
 the map~\eqref{eq3.3}  results in a  first order discretization of  \eqref{lin_sys},  which  is linear and of the form \eqref{dis_lin_sys}.
} 

\olive{Proof of \ref{item2} : To simplify the notation,} define $\vectortemp = \tau_\manifoldM(\discmap^{-1}(\extendedstate,\extendedstate^{+}))$. Then from \eqref{recursion_disc}, we have $(\extendedstate,\extendedstate^{+}) = \discmap(\vectortemp,h\mathbf{F}(\vectortemp,v))$ and consequently we have
\begin{equation*}
(\extdiff\times\extdiff)(\extendedstate,\extendedstate^{+}) = (\extdiff\times\extdiff) (\discmap(\vectortemp,h\mathbf{F}(\vectortemp,v)))    
\end{equation*}
Substituing $\discmap$ from \eqref{linearizingdisc} and expanding we get
\begin{equation}
\label{eq3.6}
\begin{split}
(\extdiff(\extendedstate),\extdiff(\extendedstate^{+})) &=   \discmap'(\extdiff(\vectortemp),h\D\extdiff(\vectortemp) \mathbf{F}(\vectortemp,v))\\
&= \discmap'(\extdiff(\vectortemp),h(\mathbf{A}\extdiff(\vectortemp)+\mathbf{B}v)).
\end{split}
\end{equation}
Substituting \red{$\discmap^{-1} = \T\extdiff^{-1}\circ\discmap'^{-1}\circ(\extdiff\times\extdiff)$} in the expression of $\vectortemp$ we have
\begin{equation*}
\begin{split}
\extdiff(\vectortemp) &= \extdiff\big(\tau_\manifoldM\big(\discmap^{-1}(\extendedstate,\extendedstate^{+}))\big)\\
&=\extdiff\big(\tau_\manifoldM\circ\T\extdiff^{-1}\circ\discmap'^{-1}\circ(\extdiff\times\extdiff)(\extendedstate,\extendedstate^+) \big)\\
&=\extdiff\big(\tau_\manifoldM\circ\T\extdiff^{-1}\big(\discmap'^{-1}(\extdiff(\extendedstate),\extdiff(\extendedstate^{+
})\big)\big)\\
\end{split}   
\end{equation*}
For any $(z,\rho)\in\T\manifoldN$, we have $\tau_\manifoldM\circ\T\extdiff^{-1}(z,\rho) = \tau_\manifoldN\big(\extdiff^{-1}(z),D\extdiff^{-1}(z).\rho\big) = \extdiff^{-1}(z) = \extdiff^{-1}\circ\tau_\manifoldN(z,\rho)$. Therefore we have
\begin{equation*}
\begin{split}
 \extdiff(\vectortemp)&= \extdiff\big(\extdiff^{-1}\circ\tau_\manifoldN(\discmap'^{-1}\big(\extdiff(\extendedstate),\extdiff(\extendedstate^{+}))\big)\big)\\
 & = \tau_\manifoldN\big(\discmap'^{-1}(\extdiff(\extendedstate),\extdiff(\extendedstate^{+}))\big) \eqqcolon \vectortemp'
\end{split}    
\end{equation*}
Substituting for $\extdiff(\vectortemp)$ in \eqref{eq3.6}, results in 
$\discmap'^{-1}(\extdiff(\extendedstate),\extdiff(\extendedstate^{+})) =(\vectortemp',h(\mathbf{A}\vectortemp'+\mathbf{B}v))$
which using \eqref{eq3.3}, further implies,
\begin{equation*}
\extdiff(\extendedstate^{+}) = \mathbf{A}_h\extdiff(\extendedstate)+\mathbf{B}_hv=\extdiff(\mathbf{F_h}(\extendedstate,v)).     
\end{equation*}
Setting $\extdiff(\extendedstate) =z$, we show \eqref{dis_ext_sys} is diffeomorphic to a linear system \eqref{dis_lin_sys} and hence is flat.
\end{proof}
\begin{rmk}
Observe that
\begin{enumerate}[label=\textup{(\roman*)},leftmargin=*, widest=b, align=left]
    \item 
    \olive{The discretization map} $\discmap$, as defined in \eqref{linearizingdisc}, also preserves the linearizing coordinate transformation $\extdiff$, and \olive{consequently,} the flat outputs.
 \item For each choice of $\discmap'$, one gets a different  $\discmap$ and \maroon{consequently}, one gets different discretization schemes of type \eqref{dis_ext_sys} for \eqref{preconsys} that are feedback linearizable.     
    \item Even though the control input $v(t)$ for the extended system \eqref{preconsys} can be kept constant on the interval $[kh,(k+1)h[,$ an exact implementation of the feedback $u(t) = \psi(x(t),y(t),v(t))$ may not be constant over $[kh,(k+1)h[$. To overcome this, we implement a zero order hold control with control held at $u(t) = u[k]\coloneqq \psi(x[k],y[k],v[k])$ for all $t\in[kh,(k+1)h[$. This limits the order of accuracy of the discretization scheme. Irrespective of the order of accuracy of \eqref{dis_lin_sys}, \maroon{the accuracy of  \eqref{dis_ext_sys} can, in general, be guaranteed upto first order.}


\end{enumerate}
\end{rmk}
\maroon{Fig. \ref{fig:my_label2} shows a schematic of discrete time implementation of a flat system.}
\section{An Illustrative Example}
In order to \olive{illustrate Theorem~\ref{main_result},}
consider the following dynamical system 
\begin{equation}
\label{ex1}
\pmat{\dot x_1\\\dot x_2\\\dot x_3\\\dot x_4} = 
\pmat{x_2+2x_2x_3\\x_3\\0\\0}+\pmat{0\\0\\1\\0}u_1+\pmat{2x_2x_4\\x_4\\0\\1+x_3}u_2,
\end{equation}
\brown{around any $x_0\in \R[4]$ \maroon{and $u_0\in \R[2]$} such that 
$x_0 \neq (\cdot, 0, -1, 0)$} \maroon{and $1+x_{0_{3}} - u_{0_2} x_{0_{4}} \neq 0$}.
\maroon{We define the dynamic compensator in two steps: first, we introduce the prolongation of control $u_2$ by setting $y = u_2, \dot{y}= \olive{\mu_2, u_1 = \mu_1}$, 
resulting in the extended system
\begin{equation*}
\pmat{\dot x_1\\\dot x_2\\\dot x_3\\\dot x_4\\\dot y} = 
\pmat{x_2+2x_2x_3\\x_3\\0\\0\\0}+\pmat{2x_2x_4\\x_4\\0\\1+x_3\\0}y+\pmat{0\\0\\
\olive \mu_1\\0\\0}+\pmat{0\\0\\0\\0\\ \olive  \mu_2},
\end{equation*}
and then \olive{introduce the following local invertible feedback
\begin{align*}
\pmat{\mu_1\\\mu_2} = \psi (x,y,v) &= \pmat{1&x_4\\ y& 1+x_3}^{-1}\pmat{v_1-(1+x_3)y^2\\v_2}\\&= 
\pmat{\frac{(1+x_3)v_1-x_4v_2-(1+x_3)^2y^2}{1+x_3-x_4y}\\\frac{yv_1+v_2-(1+x_3)y^3}{1+x_3-x_4y}},
\end{align*}
where $v:=(v_1,v_2)$ is the new control. 
}}
Then the precompensated system is given by:  

\remove{Consider a dynamic precompensator $\dot{y}=\extendedcontrol_1(x,y,v)$  and a feedback $(u_1,u_2)=\psi(x,y,v) = (\extendedcontrol_2(x,y,v),y)$, with  
with $v\coloneqq(v_1,v_2)$ and
\begin{equation*}
\pmat{\extendedcontrol_1(x,y,v)\\\extendedcontrol_2(x,y,v)} = \pmat{1&x_4\\ y& 1+x_3}^{-1}\pmat{v_1-(1+x_3)y^2\\v_2}
\end{equation*}}
{\begin{equation}
\label{ex1_comp}
\dot{\extendedstate}\coloneqq\pmat{\dot x_1\\\dot x_2\\\dot x_3\\\dot x_4\\\dot{y}} = 
\pmat{x_2+2x_2(x_3+x_4y)\\x_3+x_4y\\ \olive \psi_1(x,y,v)\\(1+x_3)y\\ \olive \psi_2(x,y,v)}\eqqcolon\mathbf{F}(\extendedstate,v),
\end{equation}
with $\extendedstate\coloneqq(x,y)$,
and \olive{can be transformed, locally around $(x_0, y_0)$ such that $1+x_{0_{3}} - y_0 x_{0_{4}} \neq 0$,   via the change of coordinates} $z\coloneqq(z_1,z_2,z_3,z_4,z_5)\coloneqq \extdiff(x_1,x_2,x_3,x_4,y)\\
\coloneqq (x_1-x_2^2,x_2,x_3+x_4y,x_4,(1+x_3)y)
$
into the following controllable linear system
\begin{equation}
\label{ex1_lin}
\begin{split}
&\dot{z}_1=z_2,\quad\quad\quad \dot{z_4}=z_5,\\ 
&\dot{z}_2=z_3,\quad\quad\quad \dot{z_5}=v_2.\\
&\dot{z}_3=v_1,
\end{split}        
\end{equation}
which we can compactly write as $\dot{z} = \mathbf{A} z+\mathbf{B}v$. Thus, 
\eqref{ex1} is linearizable by a dynamic endogenous invertible feedback and therefore flat. Further, from \eqref{ex1_lin}, the flat outputs are $(\varphi_1,\varphi_2) = (z_1,z_4) = (x_1-x_2^2,x_4)$. 
\remove{\florentina{The red part is really to clear. Go back to the version of 2024 explain that in order to simplify he understanding, we define the precompensated system in 2 steps rather than one, as in Section II. Below what I found in my old files: 
\\
*********
\\
If one instead considers the following \cyan{dynamic} precompensator
\begin{equation}
\begin{split}
u_1 &= \extendedcontrol_1\\
u_2 &= w\\
\dot{w}& = \extendedcontrol_2\\
\end{split}
\end{equation}
and the extended system dynamics given by
\begin{equation}
\begin{split}
\label{unicycle_extended}
\pmat{\dot x_1\\\dot x_2\\\dot x_3\\\dot x_4\\\dot{w}} &= 
\pmat{x_2+2x_2(x_3+x_4w)\\x_3+x_4w\\0\\(1+x_3)w\\0}+\pmat{0\\0\\1\\0\\0}\extendedcontrol_1+\pmat{0\\0\\0\\0\\1}\extendedcontrol_2,
\end{split}
\end{equation}
then one can see that under the coordinate transformation \brown{(locally invertible around any $(x_0, w_0)$ such that $1+x_{0_{3}} - w_0 x_{0_{4}} \neq 0$)}
\begin{multline*}
(z_1,z_2,z_3,z_4,z_5)\coloneqq \extdiff(x_1,x_2,x_3,x_4,w)\\
\coloneqq (x_1-x_2^2,x_2,x_3+x_4w,x_4,(1+x_3)w)
\end{multline*}
and \brown{the invertible} static feedback
\begin{equation*}
\pmat{\extendedcontrol_1\\\extendedcontrol_2}\coloneqq \pmat{1&x_4\\ w& 1+x_3}^{-1}\pmat{v_1-(1+x_3)w^2\\v_2},
\end{equation*}
where $ v \coloneqq (v_1,v_2)$ is the modified control input,~\eqref{unicycle_extended} is equivalent to the following LTI system
\begin{equation}
\label{unicycle_lin}
\begin{split}
\dot{z_1}&=z_2\\
\dot{z_2}&=z_3\\
\dot{z_3}&=v_1\\
\dot{z_4}&=z_5\\
\dot{z_5}&=v_2.
\end{split}
\end{equation}
\maroon{Therefore} \brown{the original system~\eqref{unicycle} \maroon{is} dynamic feedback linearizable around $(x_0, u_0)$.}
\\
************
}}
\subsection{Flatness preserving discretization} 
We now use Theorem \ref{main_result} to construct a flatness preserving discretization \olive{scheme} for \eqref{ex1}. Let $\manifoldM$ be the domain on which $\extdiff$ is defined and $\manifoldN= \extdiff(\manifoldM)$ be its image. Consider the following discretization map on $\T\manifoldN$, \red{$\T\manifoldN\ni(z,\rho)\mapsto\discmap'(z,\rho) = (z,z+\rho)\in\manifoldN\times\manifoldN$}, resulting in the discretization scheme: $\discmap'^{-1}(z,z^+) = (\discmap'^{-1}(z,z^+),h(\mathbf{A}\discmap'^{-1}(z,z^+)+
\mathbf{B}v)$ for \eqref{ex1_lin} resulting in the linear discrete system
\begin{equation}
\label{ex1_lin_disc}
\begin{split}
\linstate_1^+ &= \linstate_1+h\linstate_2, \quad\quad\quad \linstate^+_{4}=\linstate_4+h\linstate_{5},\\
\linstate^+_{2} &= \linstate_{2}+h\linstate_{3},\quad\quad\quad \linstate^+_{5}=\linstate_{5}+hv_{2}.\\
\linstate^+_{3} &= \linstate_{3}+hv_{1},
\end{split}
\end{equation}
which we denote compactly as $z^+ = \mathbf{A}_hz+\mathbf{B}_hv$. Then, 
\begin{equation*}
\begin{split}
(\extendedstate,\nu)\mapsto \discmap(\extendedstate,\nu) &\coloneqq \big((\extdiff^{-1}\times\extdiff^{-1})\circ \discmap' \circ \T\extdiff\big)(\extendedstate,\nu)\\
&\coloneqq \big(\extendedstate,\extdiff^{-1}(\extdiff(\extendedstate)+\mathrm{D}\extdiff(\extendedstate)\cdot\nu)\big),
\end{split}
\end{equation*}
is a discretization map on $\T\manifoldM$ and results in the following discretization scheme of \eqref{ex1_comp}: 
\begin{equation*}
\pmat{x^+\\y^+}\coloneqq\extendedstate^+ = \extdiff^{-1}(\extdiff(\extendedstate)+h\D\extdiff(\extendedstate)\mathbf{F}(\extendedstate,v))\eqqcolon\pmat{f_h(x,y,v)\\g_h(x,y,v)}
\end{equation*}
which is diffeomorphic to \olive{the} linear discrete system \eqref{ex1_lin_disc} by a coordinate change $\extdiff$, and thus, is flat with discrete time flat outputs $(\varphi_1,\varphi_2)=(z_1,z_4) \coloneqq (x_1-x_2^2,x_4)$. The corresponding continuous time control input $u$ for \eqref{ex1} is obtained using zero order hold and is given by $$u(t) = u[k] \coloneqq (\psi(x[k],y[k],v[k]),y[k]),~ t\in[kh,(k+1)h[.$$
\subsection{Trajectory tracking using difference flatness}
We utilize flatness to track given reference \olive{trajectories 
$t\mapsto \varphi^*_i(t)$, $i\in\{1,2\}$, 
for the flat output components of \eqref{ex1}.} The discretized reference trajectories are obtained by sampling \olive{the continuous time ones}, and are given by $k\mapsto \varphi_i^*[k]= \varphi_i^*(kh)$, where $h$ is the stepsize. Since \eqref{ex1_lin_disc} is flat, the reference state trajectories $k\mapsto z_i^*[k]$ and the nominal input $k\mapsto v^*[k]$ can be uniquely expressed as functions of flat outputs and their forward shifts. Let $k\mapsto e[k] = z[k]-z^*[k]$ be the tracking error. Consider a control input $v[k] = v^*[k]+Ke[k]$, where $K\in\R[2\times5]$ is \red{the} gain vector such that $\mathbf{A}_h+\mathbf{B}_hK$ has eigenvalues inside the unit circle, then $z[k]\lra z^*[k]$ as $k\lra\infty$ resulting in asymptotic tracking of the discrete time trajectories, and applying $u(t)$ as defined with zero hold results in continuous time $x(t),y(t)$ trajectories converging to $(x^*(t),y^*(t)) = \extdiff^{-1}(z^*(t))$.  
\maroon{
\begin{rmk}
Since the obtained discretization is approximate, in particular, accurate only upto first order, in order to apply control $u(t)$ effectively, the integrator must be reset periodically \cite{blanes2017concise}. While, we do not formally prove it here, the numerical experiments suggest that the continuous time system trajectories converge to reference trajectories. 
\end{rmk}
}
\section{Simulation Results}
In order to demonstrate our results, we simulated \eqref{ex1} using MATLAB. The simulation horizon was set to $10$ seconds and the stepsize $h$ was set to $0.05$ seconds (50 milliseconds). The initial condition was chosen as $(x(0),\red y(0)) = ((0.5,0.2,0.1,0.2),0)$. The reference trajectories were chosen as $(z_1^*(t),z_4^*(t)) =(0.3+0.05\sin(0.4\pi t),0.1+0.05\sin(0.2\pi t))$ for all $t\in[0,10]$. The control input was set as $v[k]= v^*[k]+Ke[k]$, with $$v^* = \left(
\frac{\fwshift[3]z_1^*-3\fwshift[2]z_1^*+3\fwshift z_1^*-z_1^*}{h^3},\frac{\fwshift[2]z_4^*-2\fwshift z_4^*+z_4^*}{h^2}\right),$$ and $K= -\pmat{10&10&10&0&0\\0&0&0&10&10}$ resulting in the Eigen values of $A_h+B_hK$ as $\big((-0.73+j0),(0.12\pm j0.81),(0.25\pm j0.66)\big)$, with $j=\sqrt{-1}$. 

Since numerical integrators are approximate and accumulate errors over time, they must be reset periodically \cite{blanes2017concise}. The integrator was reset every $1$ second (20 cycles) by setting $x[k]=x(kh)$.  The simulation results are shown in Fig. \ref{fig:placeholder}
The state and control plots for the discretized system are shown in Plots (a) and (b), respectively. The relative error $\norm{x[k]-x(kh)}/\norm{x(kh)}$ where $x(kh)$ is the exact trajectory 
(obtained by ODE45 solver of MATLAB) of~\eqref{ex1} sampled at $t=kh$, is plotted in Plot (c). Finally, Plot (d) shows that $(z_1(t),z_4(t))$ asymptotically track 
$z_1^*(t),z_4^*(t)$. 
\begin{figure}
    \centering
    \includegraphics[width = \linewidth]{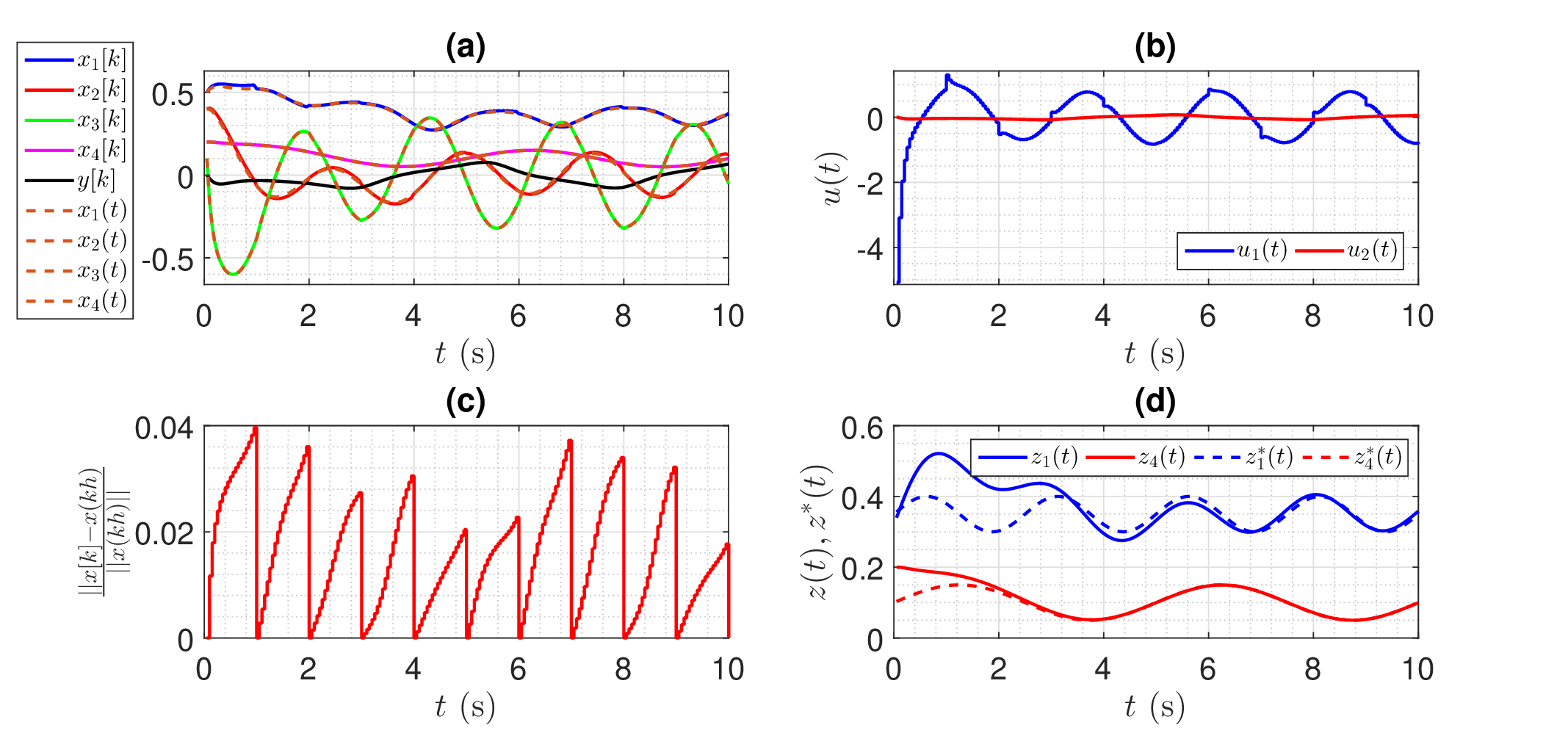}
    \vspace{-0.3in}
   \caption{Simulation Results for $h=50$ milliseconds: (a) system trajectories $x[k],x(t)$, (b) Control input $u$, (c) Relative  error $\norm{x[k]-x(kh)}/\norm{x(kh)}$, the accumulated error is reset every 1 second by setting $x[k]=x(kh)$ and  (d) Flat outputs $(z_1,z_4)$ tracking reference $(z^*_1,z^*_4)$}
    \label{fig:placeholder}
\end{figure}
\section{Conclusion}
In this article, we have extended the results of \cite{retr_disc_map} to flat systems \olive{and provided a family of discretization schemes that preserve flatness.}
Since the class of differently (and difference) flat systems is equivalent to the class of systems linearizable by dynamical endogenous feedback, Theorem~\ref{main_result} allows us to construct,   \olive{for that class of systems, discretization schemes leading to discrete systems that are feedback linearizable.} 
One of the key features of the presented results is that,
\olive{except differently flatness, we}
do not assume any specific form for the dynamics of the continuous time system. \olive{Therefore, our results apply to 
to a large class of systems (in particular, to all nonlinear control systems known to be flat)}. Moreover, such discretization schemes also preserve the flat outputs, facilitating easy implementation in discrete time.




\remove{\section{Reviewers Comments}
\subsection{Reviewer2773 -- Positive}
Main comments: The work is well written, offering a good structure and
comprehensibility. Please consider the following remarks:

\begin{enumerate}
    \item It is unclear if there are other existing methods for
this problem. --\blue{to the best of the authors' knowledge, no constructive method ensuring preservation of feedback linearization exists in literature.}
\item Since sampling is ubiquitous in practice, the question
arises to what the advantages are compared to competing
approaches -- \blue{the method preserves feedback linearization, generating controls in a discrete time setting is much simpler than arbitrary discretization, however, once a control scheme is decided, convergence of the resulting numerical integrator is to be verified. }
\item Under (A. Contributions) mentioned "very few
assumptions" are not clearly stated.
--\blue{redefine assumptions}
\item Readability and labeling of the plots could be
improved. Variables are not mentioned in the simulation and
results section.
 -- \blue{Plots should be redrawn...resimulate?}
\item Trajectories of x and u of the original system seem
like a good supplement to see that the results are
consistent. -- \blue{redraw trajectries}
\item The claim of the methods applicability to endogenous
and exogenous types of feedback is not sufficiently
explained. -- \blue{explain/discuss endogenous and exogenous feedback}
\end{enumerate}
\subsection{Reviewer 277 -- mixed}
Main comments -- This paper is an extension of previous work [14] by same
authors, on the interesting topic of preserving feedback
linearizability after discretization: the case of static
feedback linearization considered in [14] is here extended
to dynamic feedback linearization. This looks fine.
However, since the definition  of dynamic feedback
linearization is given in terms of static feedback
linearization for an extended system (def 2.3), it is not
completely clear in which sense the problem is different
from that of [14] ? In addition, even though beyond the
strict scope of the paper, nothing is said about conditions
for dynamic feedback linearizability (while this is
commented for the case of static feedback linearizability).
Finally, the discussion about retraction and discretization
maps looks very formal, and its practical use could be
discussed a bit more. -- \blue{reiterate the key difference between dynamic and static feedback linearization, or iterate that it is a demonstration of the extent of applicability of results of [14] in the context of dynamic feedback linearization and differential flatness}
\subsection{Reviewer 2779 -- }
Main comments -- 
\begin{enumerate}
    \item I have a technical concern related to applications. As I
understand it, the system itself should be discretized in
the same way as the control for the results to hold. In
practice, the system would remain continuous. I would
suggest that the authors clarify this in the revision. The
way I now see it is that you are proposing an integration
technique for nonlinear manifolds. This doesn’t relate to
how the discretization of the controller needs to be
executed. Since systems (2.11) and system (2.10) are
equivalent, one can simply construct a discrete controller,
$v_k$, that stabilizes (2.10) and be done with it provided
that we can send $\alpha(\xi(t)) + \beta(\xi(t))*v_k$ to the
actual system. The real problem with discretizing dynamic
feedback linearization is how to find a constant $\mu_k$ over
$[t_k,t_{k+1})$ that stabilizes (2.11). I don’t see how the
approach of the authors solves this problem. -- \blue{the control is to be designed post discretization, using the results from discrete LTI systems and then utilize the feedback linearization to pull back the control to obtain an $u_k$ and then correspondingly $u(t)$.}
\item I belief that the presentation of the work can be
improved considerably by incorporating why discretization
does impede the static (SFL) or dynamic feedback
linearizability (DFL). Now the authors refer the reader to
the literature whereas I consider this a key aspect to
include. Instead the authors choose to spend space to
explaining SFL/DFL itself whereas I would presume that any
reader that consults this paper would be familiar with
these concepts. A brief definition of DFL should do to
establish the notation system.
\item Given that it is critically important to this paper’s
motif, I would use the space gained by implementation of
the former comment, to show that (3.1) is in fact no longer
DFL as is apparently a consequence of [21, Theorem 3.1].
\item In proposition 3.5 a discretization of vector fields is
proposed. I have several questions related to this
definition.  -- \red{Redefine this notion}
\item The argument of the right-hand side F is confusing.
Since $\tau_X$ is defined as the canonical projection and
$D^{-1}$ as the inverse discretization map which is not
defined but I assume it produces $(x,\dot{x})$. Hence, it
reads as $F(x_k)$? Why not simply state it as such? In the
end, I	believe, $x_{k+1}$ is the retraction of the Euler
discretization? In case I’m overlooking something, it is
not easy to deduce from the text. -- \red{redefine notions}
\end{enumerate}
\subsection{Reviewer 2781 -- positive}
Main Comments --
\begin{enumerate}
    \item DFL is connected to flatness. For discrete time systems
major progress has recently been made, regarding the
definition of flatness, the test of (forward)flatness, and
also the topic of flatness preserving discretization has
been addressed, see the work of Millérioux, Kaldmäe, Kolar,
Diwold,.... -- \blue{read literature.}
\end{enumerate}
\subsection{Reviewer 2787 -- negative}
Main Comments -- 
\begin{enumerate}
    \item You suggest to calculate a discretized system from a
linearized continuous time system to meet a discrete
system, approximating to the first order, that is again
linearizable by state feedback. The reason why this would
be relevant is not made clear. Indeed, in practice one
discretizes the control law. This is not mentioned and no
related reference is given. -- \red{The reviewers questions the need for feedback linearizability-preserving discretization itself.}

After extending the system by a dynamic extension you just
applied what you have published before. This is trivial.
So, I do not see any (significant) contribution. -- \red{Indeed that is the case, the paper primarily deals with the extent of applicability of results of [14] in the dynamic fl context}

You consider flat systems and speak about exogenous and
endogenous feedback,
which are concepts introduced in the flatness context, but
you never mention flatness. The references (7-9) you cite
are quite irrelevant, and it is not made clear why you
choose these latter ones. -- \red{introduce flatness}
\item What is the difference of your reasoning about first order
discretization 
and just speaking about locally linearized systems.
The relevant facts are not properly explained while many
repetitions need much space. 

Even the discussion of the simulation example is weak. What
is the meaning of seconds? What are the eigenvalues chosen?
Plotting absolute errors as in Fig. 3 is meaningless as it
is arbitrarily scalable.
\end{enumerate}}
\bibliographystyle{IEEEtran}
\bibliography{IEEEabrv,name1}

\begin{thebibliography}{10}
\providecommand{\url}[1]{#1}
\csname url@samestyle\endcsname
\providecommand{\newblock}{\relax}
\providecommand{\bibinfo}[2]{#2}
\providecommand{\BIBentrySTDinterwordspacing}{\spaceskip=0pt\relax}
\providecommand{\BIBentryALTinterwordstretchfactor}{4}
\providecommand{\BIBentryALTinterwordspacing}{\spaceskip=\fontdimen2\font plus
\BIBentryALTinterwordstretchfactor\fontdimen3\font minus
  \fontdimen4\font\relax}
\providecommand{\BIBforeignlanguage}[2]{{%
\expandafter\ifx\csname l@#1\endcsname\relax
\typeout{** WARNING: IEEEtran.bst: No hyphenation pattern has been}%
\typeout{** loaded for the language `#1'. Using the pattern for}%
\typeout{** the default language instead.}%
\else
\language=\csname l@#1\endcsname
\fi
#2}}
\providecommand{\BIBdecl}{\relax}
\BIBdecl

\bibitem{retr_disc_map}
A.~Jindal, R.~Banavar, and D.~M. Diego, ``Constructing feedback linearizable
  discretizations for continuous-time systems using retraction maps,''
  \emph{IEEE Control Systems Letters}, vol.~7, pp. 2467--2472, 2023.

\bibitem{fliess1995flatness}
M.~Fliess, J.~L{\'e}vine, P.~Martin, and P.~Rouchon, ``Flatness and defect of
  non-linear systems: introductory theory and examples,'' \emph{International
  Journal of Control}, vol.~61, no.~6, pp. 1327--1361, 1995.

\bibitem{wang2022trajectory}
X.~Wang and W.~Sun, ``Trajectory tracking of autonomous vehicle: A differential
  flatness approach with disturbance-observer-based control,'' \emph{IEEE
  Transactions on Intelligent Vehicles}, vol.~8, no.~2, pp. 1368--1379, 2022.

\bibitem{martin:cel-00392180}
\BIBentryALTinterwordspacing
P.~Martin, P.~Rouchon, and R.~M. Murray, ``{Flat systems, equivalence and
  trajectory generation},'' Aug. 2006, lecture. [Online]. Available:
  \url{https://cel.hal.science/cel-00392180}
\BIBentrySTDinterwordspacing

\bibitem{kolar2017time}
B.~Kolar, H.~Rams, and K.~Schlacher, ``Time-optimal flatness based control of a
  gantry crane,'' \emph{Control Engineering Practice}, vol.~60, pp. 18--27,
  2017.

\bibitem{murray1995differential}
R.~M. Murray, M.~Rathinam, and W.~Sluis, ``Differential flatness of mechanical
  control systems: A catalog of prototype systems,'' in \emph{ASME
  international mechanical engineering congress and exposition}.\hskip 1em plus
  0.5em minus 0.4em\relax Citeseer, 1995.

\bibitem{mellinger2011minimum}
D.~Mellinger and V.~Kumar, ``Minimum snap trajectory generation and control for
  quadrotors,'' in \emph{2011 IEEE international conference on robotics and
  automation}.\hskip 1em plus 0.5em minus 0.4em\relax IEEE, 2011, pp.
  2520--2525.

\bibitem{greeff2018flatness}
M.~Greeff and A.~P. Schoellig, ``Flatness-based model predictive control for
  quadrotor trajectory tracking,'' in \emph{2018 IEEE/RSJ International
  Conference on Intelligent Robots and Systems (IROS)}.\hskip 1em plus 0.5em
  minus 0.4em\relax IEEE, 2018, pp. 6740--6745.

\bibitem{sun2022comparative}
S.~Sun, A.~Romero, P.~Foehn, E.~Kaufmann, and D.~Scaramuzza, ``A comparative
  study of nonlinear mpc and differential-flatness-based control for quadrotor
  agile flight,'' \emph{IEEE Transactions on Robotics}, vol.~38, no.~6, pp.
  3357--3373, 2022.

\bibitem{welde2022role}
J.~Welde, M.~D. Kvalheim, and V.~Kumar, ``The role of symmetry in constructing
  geometric flat outputs for free-flying robotic systems,'' in \emph{2023 IEEE
  International Conference on Robotics and Automation (ICRA)}.\hskip 1em plus
  0.5em minus 0.4em\relax IEEE, 2023, pp. 12\,247--12\,253.

\bibitem{kolar2022necessary}
B.~Kolar, J.~Diwold, and M.~Sch{\"o}berl, ``Necessary and sufficient conditions
  for difference flatness,'' \emph{IEEE Transactions on Automatic Control},
  vol.~68, no.~3, pp. 1715--1721, 2022.

\bibitem{levine2009analysis}
J.~L{\'e}vine, \emph{{Analysis and Control of Nonlinear Systems: A
  Flatness-Based Approach}}.\hskip 1em plus 0.5em minus 0.4em\relax Springer,
  2009.

\bibitem{martin2001part1}
P.~Martin, R.~Murray, and P.~Rouchon, ``Flat systems, equivalence and
  feedback,'' in \emph{Advances in the control of nonlinear systems},
  A.~Ba{\~{n}}os, F.~Lamnabhi-Lagarrigue, and F.~J. Montoya, Eds.\hskip 1em
  plus 0.5em minus 0.4em\relax London: Springer London, 2001, pp. 5--32.

\bibitem{martin2001part2}
------, ``Flat systems: open problems, infinite dimensional extension,
  symmetries and catalog,'' in \emph{Advances in the control of nonlinear
  systems}, A.~Ba{\~{n}}os, F.~Lamnabhi-Lagarrigue, and F.~J. Montoya,
  Eds.\hskip 1em plus 0.5em minus 0.4em\relax London: Springer London, 2001,
  pp. 33--57.

\bibitem{diwold2021trajectory}
J.~Diwold, B.~Kolar, and M.~Sch{\"o}berl, ``A trajectory-based approach to
  discrete-time flatness,'' \emph{IEEE Control Systems Letters}, vol.~6, pp.
  289--294, 2021.

\bibitem{kolar2016construction}
B.~Kolar, A.~Kaldm{\"a}e, M.~Sch{\"o}berl, {\"U}.~Kotta, and K.~Schlacher,
  ``Construction of flat outputs of nonlinear discrete-time systems in a
  geometric and an algebraic framework,'' \emph{IFAC-PapersOnLine}, vol.~49,
  no.~18, pp. 796--801, 2016.

\bibitem{diwold2023theory}
\BIBentryALTinterwordspacing
J.~Diwold, ``Theory and applications of discrete-time flatness,'' Ph.D.
  dissertation, JOHANNES KEPLER UNIVERSITY LINZ, 2023. [Online]. Available:
  \url{https://resolver.obvsg.at/urn:nbn:at:at-ubl:1-60402}
\BIBentrySTDinterwordspacing

\bibitem{diwold2022discrete}
J.~Diwold, B.~Kolar, and M.~Sch{\"o}berl, ``Discrete-time flatness-based
  control of a gantry crane,'' \emph{Control Engineering Practice}, vol. 119,
  p. 104980, 2022.

\bibitem{diwold2023discrete}
------, ``Discrete-time flatness-based controller design using an implicit
  euler-discretization,'' \emph{IFAC-PapersOnLine}, vol.~56, no.~1, pp.
  138--143, 2023.

\bibitem{21MBLDMdD}
M.~Barbero Li\~n\'an and D.~Mart\'in~de Diego, ``Extended retraction maps: a
  seed of geometric integrators,'' \emph{Found. Comput. Math.}, 2022.

\bibitem{fossas1998flatness}
E.~Fossas, J.~Franch, and A.~Palau, ``Flatness, tangent systems and flat
  outputs,'' in \emph{Proceedings of the 1998 American Control Conference. ACC
  (IEEE Cat. No. 98CH36207)}, vol.~1.\hskip 1em plus 0.5em minus 0.4em\relax
  IEEE, 1998, pp. 313--317.

\bibitem{sira2004differentially}
H.~Sira-Ram{\'\i}rez and S.~K. Agrawal, \emph{Differentially Flat
  Systems}.\hskip 1em plus 0.5em minus 0.4em\relax CRC Press, 2004.

\bibitem{martin1992phd}
P.~Martin, ``Contribution {\`a} l'{\'e}tude des syst{\`e}mes
  diff{\'e}rentiellement plats,'' Ph.D. dissertation, l'Ecole Nationale
  Sup{\'e}rieure de Mines de Paris, 1992.

\bibitem{pomet1995differential}
J.~Pomet, ``A differential geometric setting for dynamic equivalence and
  dynamic linearization,'' \emph{Banach Center Publ., Vol. 32}, pp. 319--339,
  1995.

\bibitem{l1980linearization}
B.~Jakubczyk and W.~Respondek, ``On linearization of control systems,''
  \emph{Bull. Acad. Polon. Sci. Ser. Sci. Math}, vol.~28, no. 9-10, pp.
  517--522, 1980.

\bibitem{kaldmae2013flatness}
A.~Kaldm{\"a}e \emph{et~al.}, ``On flatness of discrete-time nonlinear
  systems,'' \emph{IFAC Proceedings Volumes}, vol.~46, no.~23, pp. 588--593,
  2013.

\bibitem{guillot2019flatness}
P.~Guillot and G.~Mill{\'e}rioux, ``Flatness and submersivity of discrete-time
  dynamical systems,'' \emph{IEEE Control Systems Letters}, vol.~4, no.~2, pp.
  337--342, 2019.

\bibitem{aranda2008linearization}
E.~Aranda-Bricaire and C.~H. Moog, ``Linearization of discrete-time systems by
  exogenous dynamic feedback,'' \emph{Automatica}, vol.~44, no.~7, pp.
  1707--1717, 2008.

\bibitem{grizzle1986feedback}
J.~Grizzle, ``Feedback linearization of discrete-time systems,'' in
  \emph{Analysis and Optimization of Systems}.\hskip 1em plus 0.5em minus
  0.4em\relax Springer, 1986, pp. 273--281.

\bibitem{kottainversion}
{\"U}.~Kotta, ``Inversion method in the discrete-time nonlinear control systems
  synthesis problems,'' 1995.

\bibitem{grizzle1988feedback}
J.~Grizzle and P.~Kokotovic, ``Feedback linearization of sampled-data
  systems,'' \emph{IEEE Transactions on Automatic Control}, vol.~33, no.~9, pp.
  857--859, 1988.

\bibitem{blanes2017concise}
S.~Blanes and F.~Casas, \emph{{A} {C}oncise {I}ntroduction to {G}eometric
  {N}umerical {I}ntegration}.\hskip 1em plus 0.5em minus 0.4em\relax CRC press,
  2017.

\end{thebibliography}
\end{document}